\documentclass{PoS}

\def\spose#1{\hbox to 0pt{#1\hss}}
\def\ltapprox{\mathrel{\spose{\lower 3pt\hbox{$\mathchar"218$}}
 \raise 2.0pt\hbox{$\mathchar"13C$}}}
\def\gtapprox{\mathrel{\spose{\lower 3pt\hbox{$\mathchar"218$}}
 \raise 2.0pt\hbox{$\mathchar"13E$}}}

\title{Electric and Magnetic Screening Masses around 
the Deconfinement Transition}

\ShortTitle{Screening Masses around Deconfinement}

\author{Attilio Cucchieri$^{ab}$, \speaker{Tereza Mendes}$^a$\\
\llap{$^a$}Instituto de F\'\i sica de S\~ao Carlos, Universidade de S\~ao Paulo, \\
           Caixa Postal 369, 13560-970 S\~ao Carlos, SP, Brazil\\
\llap{$^b$}Ghent University, Department of Physics and Astronomy, \\
           Krijgslaan 281-S9, 9000 Gent, Belgium \\
E-mail: \email{attilio@ifsc.usp.br}, \email{mendes@ifsc.usp.br}}

\abstract{We report on the status of our study of gluon propagators
and screening masses around
the deconfining transition for pure SU(2) gauge theory in Landau gauge.
}

\FullConference{ The XXIX International Symposium on Lattice Field Theory - Lattice 2011\\
July 10-16, 2011\\
Squaw Valley, Lake Tahoe, California}

\begin{document}

\section{Introduction}

Debye screening of the color charge, expected at high temperature,
is signaled by screening masses/lengths that can in principle be 
obtained from the gluon propagator \cite{Gross:1980br}.
More specifically, chromoelectric (resp.\ chromomagnetic) screening
will be related to the longitudinal (resp.\ transverse) gluon
propagator computed at momenta with null temporal component (soft modes).
In particular, we expect the real-space longitudinal
propagator to fall off exponentially at long distances,
defining a (real) electric screening mass,
which can be calculated perturbatively to leading order.
Also, according to the 3d adjoint-Higgs picture for dimensional
reduction, we expect the transverse propagator to show a confining
behavior at finite temperature, in association with a nontrivial
magnetic mass (see e.g.\ \cite{Cucchieri:2001tw}).
We note that these propagators are gauge-dependent quantities,
and the (perturbative) prediction that the propagator poles should
be gauge-independent must be checked, by considering different gauges.

Even though the nonzero-$T$ behavior just described has been verified 
for various gauges and established at high temperatures down to around
twice the critical temperature $T_c$ \cite{Cucchieri:2001tw,Heller:1997nqa},
it is not clear how a screening mass would develop around $T_c$.
At the same time, lattice studies of Landau-gauge gluon propagators at 
finite temperature in pure $SU(2)$ and $SU(3)$ theory have observed a 
sharp peak in the infrared value of the electric propagator around the 
deconfinement temperature, suggesting an alternative order parameter 
for the QCD phase transition 
\cite{Cucchieri:2007ta,Fischer:2010fx,Bornyakov:2010nc,Aouane:2011fv,Maas:2011ez}. 
(Of course, a relevant question is, then, whether this singularity survives 
the inclusion of dynamical quarks in the theory \cite{Bornyakov:2011jm}.)
In the following, we investigate the critical behavior of electric and 
magnetic gluon propagators and try to characterize the screening 
masses around the transition temperature $T_c$ by performing large-lattice 
simulations in pure SU(2) gauge theory. 
In particular, we use the knowledge gained in the
study of the zero-temperature gluon propagator (see \cite{Cucchieri:2010xr}
for a review) to identify systematic effects in the infrared limit
and to define temperature-dependent masses for the region around and 
below $T_c$.
A more detailed analysis of our data will be presented
elsewhere \cite{inprep}. (Preliminary results were reported in
\cite{Cucchieri:2011ga,Cucchieri:2011di}.)


\section{Results}
\label{results}

We have considered the pure SU(2) case, with a standard Wilson action.
For our runs we employ a cold start, performing a projection on 
positive-Polyakov-loop configurations. 
Also, gauge fixing is done using stochastic overrelaxation and
the gluon dressing functions are normalized to 1 at 2 GeV. 
We take $\beta$ values in the scaling region
and lattice sizes ranging from $N_s =$ 48 to 192 and from $N_t =$ 2 to 16 
lattice points, respectively along the spatial and along the temporal 
directions.
Our (improved) procedure for determining the physical temperature $T$ 
is described in \cite{Cucchieri:2011di}.
The momentum-space expressions for the transverse and longitudinal 
gluon propagators $D_T(p)$ and $D_L(p)$ can be found e.g.\ 
in \cite{Cucchieri:2007ta}.
We first describe our investigation of the critical behavior for
transverse and longitudinal gluon propagators and then discuss our 
proposal for computing screening masses around $T_c$.

\begin{figure}
\centerline{
\includegraphics[height=10truecm]{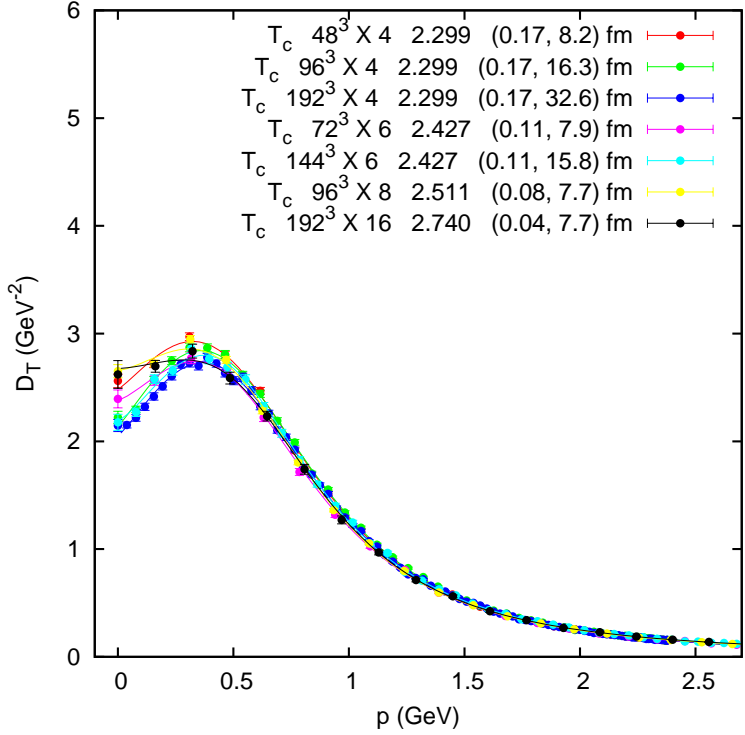}
}
\caption{Transverse gluon propagator at
$T_c$, for various choices of lattice size and $\beta$.
Values for the temperature, $\,N_s^3 \times N_t$, $\,\beta$, lattice 
spacing $a$ and spatial lattice size $L$ (both in fm, in parentheses) 
are given in the plot labels. The solid lines are fits, described at
the end of this section.}
\label{DTatTc}
\end{figure}

Our data for the transverse (magnetic) propagator $D_T(p)$
at the critical temperature are shown in Fig.\ \ref{DTatTc}.
We clearly see the strong infrared suppression of the propagator, 
as expected, with a turnover at around 400 MeV. Regarding systematic 
errors, there are considerable finite-physical-size effects, evidenced 
by the fact that the lattices with the smaller physical spatial size 
(the red, yellow, magenta and black curves) show different 
behavior from the remaining curves (green, blue and cyan).
The nature of these effects is similar to what is observed at $T=0$
\cite{Cucchieri:2010xr}.
We mention that essentially the same features are seen for $D_T(p)$
at other temperatures, above and below $T_c$ \cite{Cucchieri:2011di}.

\begin{figure}
\centerline{
\includegraphics[height=10truecm]{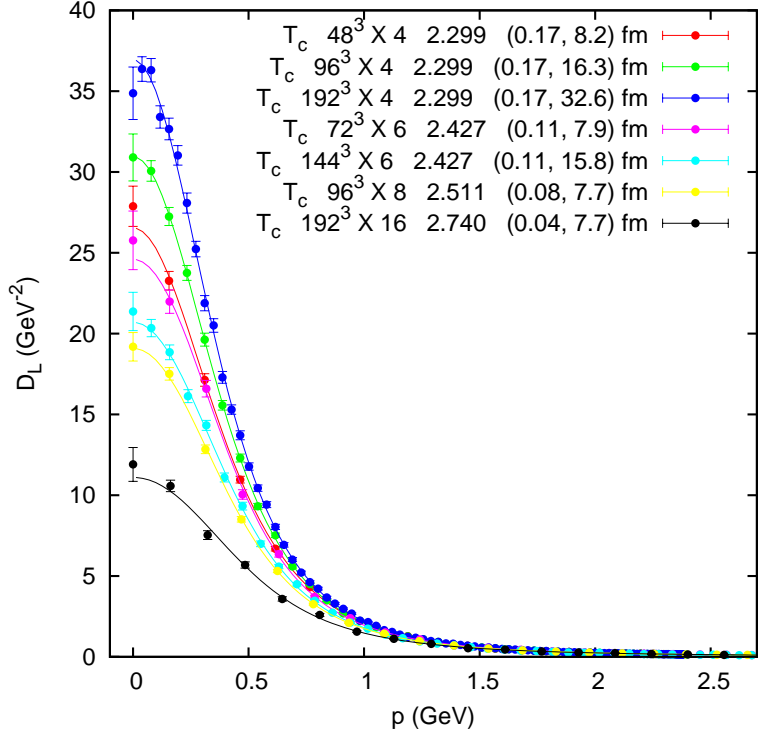}
}
\caption{Longitudinal gluon propagator at and around
$T_c$, for various choices of lattice size and $\beta$.
Values for the temperature, $\,N_s^3 \times N_t$, $\,\beta$, lattice 
spacing $a$ and spatial lattice size $L$ (both in fm, in parentheses) 
are given in the plot labels. The solid lines are fits, described at
the end of this section.}
\label{DLatTc}
\end{figure}

The longitudinal (electric) propagator $D_L(p)$ at $T_c$ is shown in Fig.\ 
\ref{DLatTc}. We immediately see severe systematic effects for the
smaller values of $N_t$. Let us note that our runs were initially planned 
under the assumption that a temporal extent $N_t = 4$ might be sufficient 
to observe the infrared behavior of the propagators and our goal was,
then, to increase $N_s$ significantly, to check for finite-size effects.
As seen in Fig.\ \ref{DLatTc}, this assumption is {\em not} justified
for the longitudinal propagator around the critical temperature, 
especially in the case of larger $N_s$.
Indeed, as $N_s$ is doubled from 48 to 96 and then to 192, we see that
the infrared value of $D_L(p)$ changes drastically, resulting in a 
qualitatively different curve at $N_s=192$, apparently with a 
turnover in momentum. (Note that, in this case, the real-space 
longitudinal propagator manifestly violates reflection positivity.)
We took this as an indication that our choice of $N_t=4$ was not 
valid and therefore considered larger values of $N_t$. 
We assume here that data points at $N_t = 16$ are essentially free from 
systematic effects, since (as shown in \cite{Cucchieri:2011di}) the 
curves for temperatures around $T_c$ stabilize for $N_t > 8$.
As seen in the figure, we obtain in this 
way a different picture for the critical behavior of $D_L(p)$. 
It is interesting to note (see Fig.\ \ref{DLatTc}) that the $N_s$ effects 
at $T_c$ are significant for $N_t=6$ (with opposite sign with respect to 
the $N_t=4$ case) and are still present for $N_t = 8$ 
(and maybe also for $N_t = 16$).
This is also true slightly below $T_c$, but not immediately above $T_c$
\cite{Cucchieri:2011di}.


In summary, the transverse propagator $D_T(p)$ shows significant
finite-physical-size effects at $T_c$, while the longitudinal propagator 
$D_L(p)$ is subject to two sources of systematic errors for small $N_t$:
``pure'' small-$N_t$ effects (associated with discretization errors) 
and strong dependence on the spatial lattice size $N_s$ at fixed $N_t$, 
when this value of $N_t$ is smaller than 16.
The latter effect was observed only at $T \ltapprox T_c$, whereas the
former is present in a wider range of temperatures around $T_c$ (see below).
For all investigated values of the temperature, $\,D_L(p)$ seems to 
reach a plateau at small momentum $p$, while $D_T(p)$ is infrared-suppressed,
with a turnover in momentum roughly around 350 MeV for all $T\neq 0$.

\begin{figure}
\hspace*{-1.5cm}
\includegraphics[height=7.2truecm]{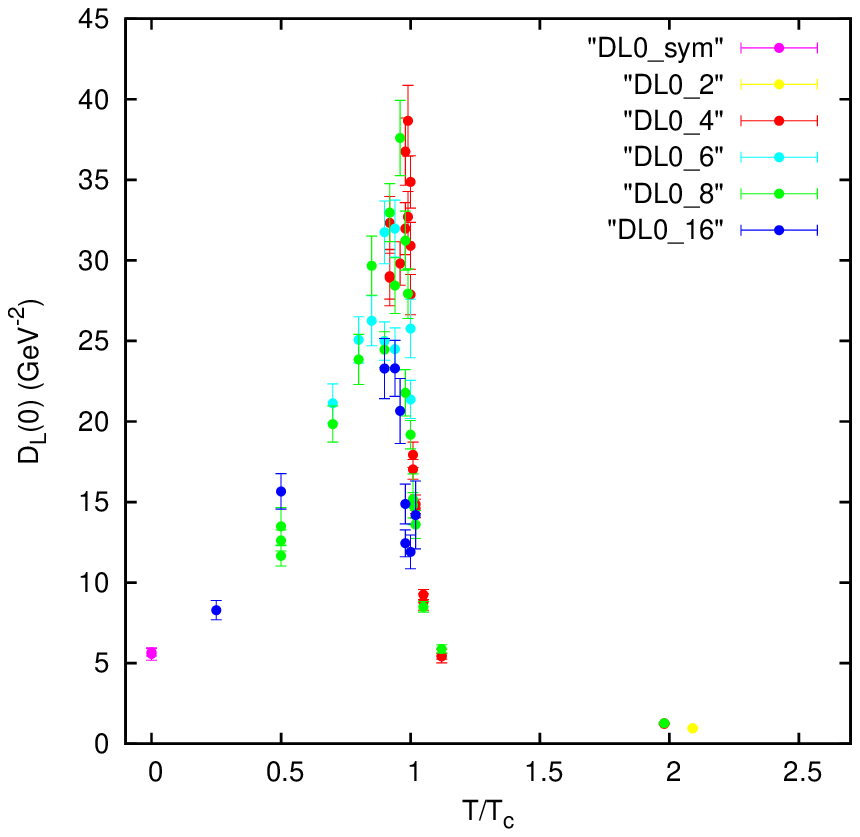}
\hspace*{-2.8cm}
\includegraphics[height=7.2truecm]{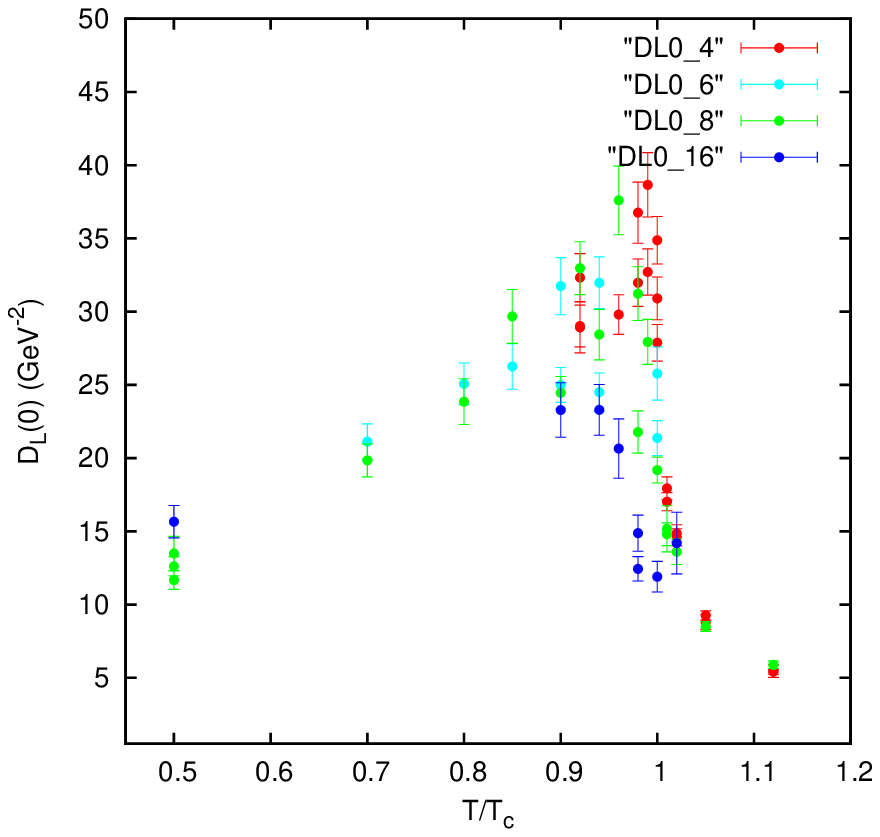}
\caption{Infrared-plateau value for the longitudinal gluon propagator
[estimated by $D_L(0)$] as a function of the temperature for the
full range of $T/T_c$ values (left) and for the region around $T_c$
(right). 
Data points from runs at the same value of $N_t$ are grouped together
and indicated by the label ``DL0\_$N_t$'', where ``sym'' is used to
indicate symmetric lattices (i.e.\ $T=0$).
}
\label{plateau}
\end{figure}

Considering the infrared plateau in $\,D_L(p)$ --- which we estimate
here by $D_L(0)$ --- as a function of temperature, the value observed 
at $\,T=0$ increases as the temperature is switched on, drops 
significantly for $T\gtapprox T_c$ and then shows a steady decrease. 
In Fig.\ \ref{plateau}, we show data for $D_L(0)$ for all our runs
on the left-hand side, and for the region around $T_c$ on the right.
We group together results from runs using the same value of $N_t$,
and indicate them by the label ``DL0\_$N_t$''.
The data points indicated with ``sym'' correspond to symmetric lattices,
i.e.\ to the zero-temperature case.
Note that results for different $N_s$'s at fixed $N_t$ may not fall on 
top of each other, which gives us an indication of the systematic errors 
discussed above. 
These are especially serious for $N_t=4$ around $T_c$ (red points).
We see that, surprisingly, the maximum value of $D_L(0)$ is not
attained for $T=T_c$ --- as might have appeared to be the case
from the $N_t=4$ lattices only --- and it does not correspond to a flat 
curve from 0.5 $T_c$ to $T_c$, as could be expected by
looking only at these two temperatures. Rather, the maximum
seems to lie at about 0.9 $T_c$. 
Moreover, it clearly corresponds to a finite peak, which does not
turn into a divergence as $N_s$ is increased at fixed $N_t$.

\begin{figure}
\hspace*{-1.5cm}
\includegraphics[height=7.1truecm]{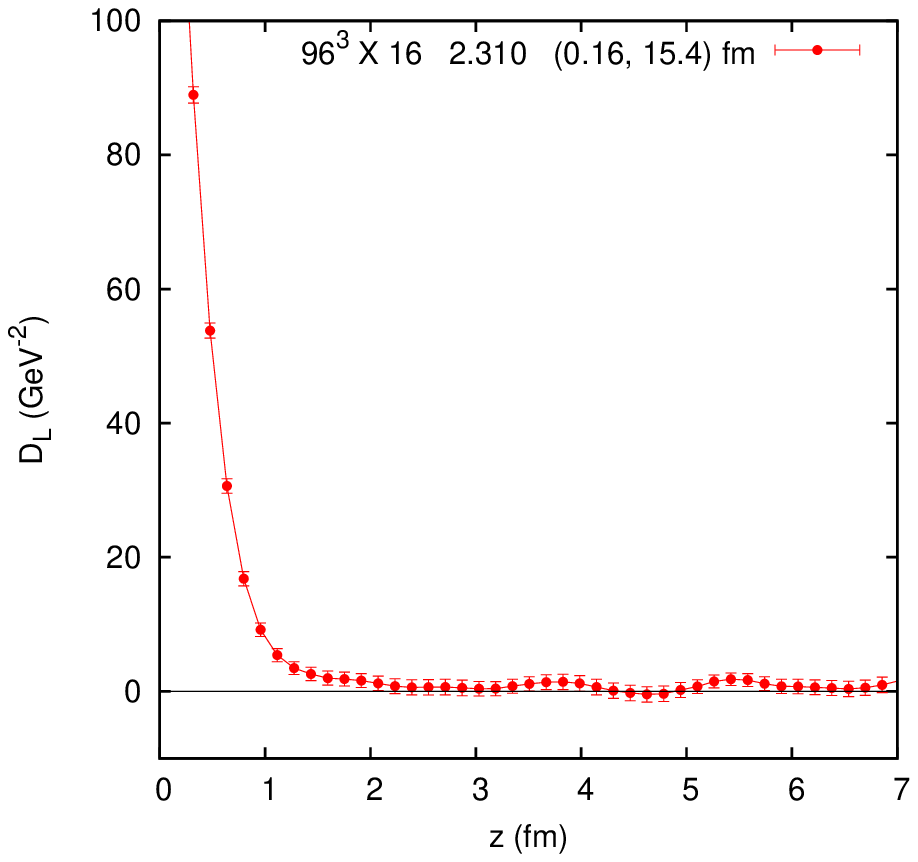}
\hspace*{-2.6cm}
\includegraphics[height=7.1truecm]{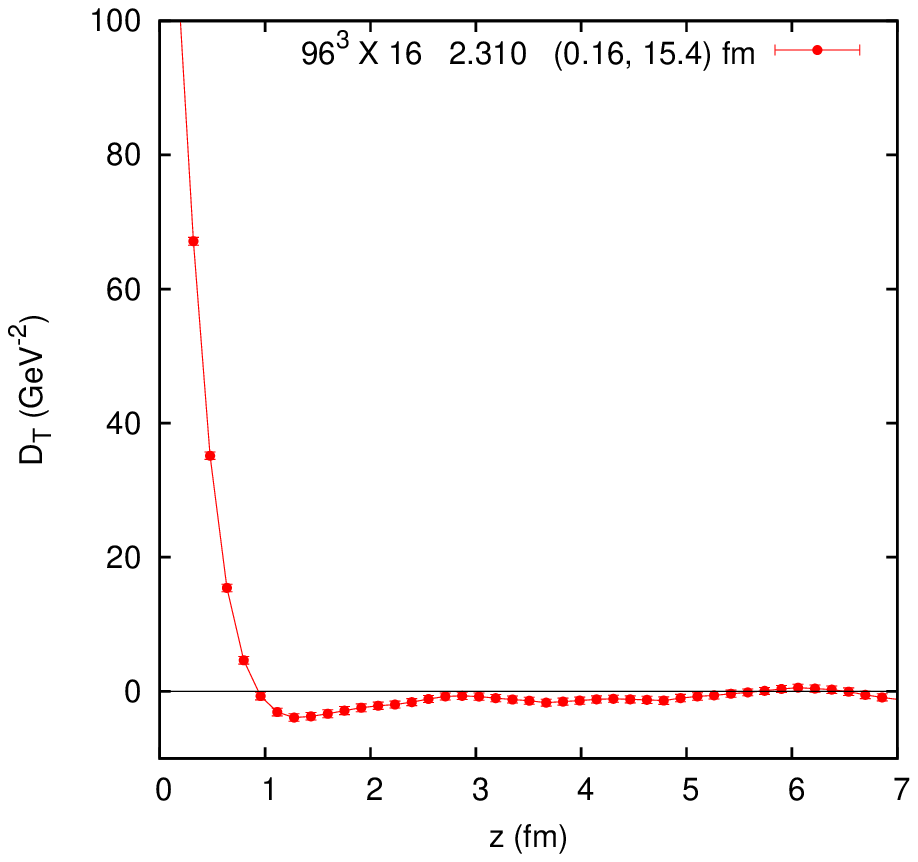}
\caption{Longitudinal (left) and transverse (right) gluon propagator
in real space for $T = 0.25 T_c$.
Values for $\,N_s^3 \times N_t$, $\,\beta$, lattice spacing $a$ and spatial
lattice size $L$ (both in fm, in parentheses) are given in the plot
labels. Note that the solid lines are {\em not} fits.}
\label{zgluon}
\end{figure}

We have also looked at the real-space propagators.
We find clear violation of reflection positivity for the transverse
propagator at all temperatures. For the longitudinal propagator,
positivity violation is observed unequivocally only at zero temperature 
and for a few cases around the critical region, in association with 
the severe systematic errors discussed above. For all other cases,
there is no violation within errors. Also, we always observe an
oscillatory behavior, indicative of a complex-mass pole.
Typical curves for the longitudinal and transverse propagators
in real space are shown (for $T=0.25T_c$) in Fig.\ \ref{zgluon}.


\vskip 3mm
We now address the problem of characterizing the screening masses
around $T_c$. For all fits shown in Figs.\ \ref{DTatTc} and \ref{DLatTc}
we used the same expression, namely a five-parameter fitting form
of the Gribov-Stingl \cite{Stingl:1985hx,Stingl:1994nk}
type\footnote{Note that, for given values of $a$, $b$, $d$, $\eta$, 
the global constant $C$ is fixed by the renormalization condition, 
so that there are only four free parameters in (\ref{GSform}).}
\begin{equation}
D_{L,T}(p) \;=\; 
C\,\frac{1\,+\,d\,p^{2 \eta}}
{(p^2 + a)^2 \,+\, b^2}\,.
\label{GSform}
\end{equation}
This form allows for two (complex-conjugate) poles, with
masses $\,m^2 \;=\; a\,\pm\, i b$, where $m \;=\; m_R \,+\, i m_I$.
The mass $m$ thus depends only on $a$, $b$ and not on the normalization $C$.
The parameter $\eta$ should be 1 if the fitting form also describes
the large-momenta region (from our infrared data we get $\eta\neq 1$).
Recall that at high temperatures one usually defines the electric screening 
mass as the scale determining the exponential decrease of the real-space 
propagator at large distances, which is equivalent to $\,D_L(0)^{-1/2}$
in the case of a real pole.
We therefore expect to observe $\,m_I\to 0\,$ (i.e.\ $\,b\to 0\,$)
for the longitudinal gluon propagator at high temperature.
Note that, if the propagator has the above form, then
the screening mass defined by $\,D_L(0)^{-1/2} \,=\, \sqrt{(a^2+b^2)/C}\;\,$
mixes the complex and imaginary masses $\,m_R$ and $m_I\,$ and depends 
on the (a priori arbitrary) normalization $C$.

We generally find good fits to the Gribov-Stingl form (including the full 
range of momenta), with 
nonzero real and imaginary parts of the pole masses in all cases.
For the transverse propagator $D_T(p)$, the masses $m_R$ and $m_I$ 
are of comparable size (around 0.6 and 0.4 GeV respectively). 
The same holds for $D_L(p)$, but in this case the relative size of 
the imaginary mass seems to decrease with increasing temperature. 
A detailed discussion of the associated masses $m_R$, $m_I$ is
postponed to a forthcoming study \cite{inprep}, as we are presently
considering variants of the above fitting form inspired by
zero-temperature studies.
Indeed, the use of a Gribov-Stingl form is motivated by the behavior of the 
gluon propagator at $\,T=0$, where this type of expression has been
shown to describe well the data in three space-time dimensions
\cite{Cucchieri:2003di}. Recently, in \cite{Cucchieri:2011ig}, various 
fitting forms of this type were used to describe large-lattice data for
the 3d and 4d gluon propagator at $\,T=0$. Noting that the 3d case
may be considered as the $T\to\infty$ limit of the 4d case, we propose to 
interpolate the 3d and 4d zero-temperature forms to describe our 
finite-$T$ data in 4d.


\section{Conclusions}

We study the longitudinal (electric) and transverse (magnetic) gluon 
propagators in momentum space, proposing the calculation of screening masses 
through an Ansatz from the zero-temperature case. Going from zero to nonzero 
temperature, we see that the electric propagator $D_L(p)$ is enhanced, with 
an apparent plateau value in the infrared, while the magnetic propagator 
$D_T(p)$ gets progressively more infrared-suppressed, with a clear turnover 
in momentum at all nonzero temperatures considered. Severe systematic 
effects are observed for the electric propagator around $T_c$, suggesting 
that lattices of temporal extent $N_t > 8$ are needed for this study. 
Once these errors are removed, the data support a finite maximum (located 
at about 0.9 $T_c$) for the infrared value of $D_L(p)$.


\section*{Acknowledgements}

The authors thank agencies FAPESP and CNPq for financial support.
A. Cucchieri also acknowledges financial support from the
Special Research Fund of Ghent University (BOF UGent).
Our simulations were performed on the new CPU/GPU cluster at
IFSC--USP (obtained through a FAPESP grant).

\end{document}